\documentclass[12pt,showpacs,amsmath,amssymb,aps,floats,floatfix,nofootinbib]{revtex4}%

\usepackage{amsmath,amssymb}
\usepackage{epsfig,lscape,rotating}
\usepackage{graphicx}
\usepackage{color}

\include{newcommands}

\def\beq{\begin{equation}}
\def\eeq#1{\label{#1}\end{equation}}
\def\eeqn{\end{equation}}
\def\beqa{\begin{eqnarray}}
\def\eeqa#1{\label{#1}\end{eqnarray}}
\def\eeqan{\end{eqnarray}}

\def\stacksymbols #1#2#3#4{\def\theguybelow{#2}
    \def\vp{\lower#3pt}
    \def\sp{\baselineskip0pt\lineskip#4pt}
    \mathrel{\mathpalette\intermediary#1}}

\def\intermediary#1#2{\vp\vbox{\sp
     \everycr={}\tabskip0pt
     \halign{$\mathsurround0pt#1\hfil##\hfil$\crcr#2\crcr
              \theguybelow\crcr}}}

\def\to{\rightarrow}

\def\x{\varphi}

\begin{document}

\title{The interpretation for Galactic Center Excess and
Electroweak Phase Transition in the NMSSM}

\author{Xiao-Jun Bi$^2$}
\author{Ligong Bian$^1$}
\author{Weicong Huang$^1$}
\author{Jing Shu$^1$}
\author{Peng-Fei Yin$^2$}

\affiliation{$^1$State Key Laboratory of Theoretical Physics, Institute of Theoretical Physics,Chinese Academy of Sciences,
Beijing 100190, China}
\affiliation{$^2$Key Laboratory of Particle Astrophysics,
Institute of High Energy Physics, Chinese Academy of Sciences,
Beijing 100049, China}

\begin{abstract}
The gamma-ray excess observed by the Fermi-LAT in the Galactic Center can be interpreted by the dark matter annihilation to $b\bar{b}$ via a light pseudoscalar in the NMSSM. It is interesting to note that the corresponding singlet scalar is useful to achieve a strongly first order phase transition required by the electroweak baryogenesis. In this paper, we investigate the possibility that the NMSSM model can simultaneously accommodate these two issues. The phase transition strength can be characterized by the vacua energy gap at zero temperature and be sufficiently enhanced by the tree-level effect in the NMSSM. We find that the annihilation of Singlino/Higgsino DM particles occurring close to the light pseudoscalar resonance is favored by the galactic center excess and the observed DM relic density, and a resulting small $\kappa/\lambda$ and a negative $A_\kappa$ can also lead to a successful strongly first order electroweak phase transition.

\end{abstract}

\maketitle

\section{Introduction}

The weakly interacting massive particles (WIMPs) are the most popular and attractive dark matter (DM) candidates. WIMPs with masses of $\mathcal{O}(1)-\mathcal{O}(10^3)$~GeV naturally provide the correct DM relic density via the freeze-out mechanism in the early Universe. The gamma rays produced by present DM annihilations are very good probe to reveal the microscopic nature and the distribution of DM due to the simple propagation and low energy loss. Such DM induced gamma-ray signatures from astrophysical sources with high DM densities, such as dwarf galaxies, galaxy clusters and the Galactic Center, have been extensively studied.

Recently, an extended gamma-ray excess of a few GeV in the GC was discovered in the Fermi Large Area Telescope data ~\cite{Goodenough:2009gk,Hooper:2010mq,Boyarsky:2010dr,Hooper:2011ti,Abazajian:2012pn,Macias:2013vya,Abazajian:2014fta,Daylan:2014rsa,Lacroix:2014eea,Zhou:2014lva,Calore:2014xka}. Although the galactic center excess (GCE) is highly statistically significant, the diffuse gamma-ray background model in the GC would introduce some unclear systematic uncertainties. After considering such uncertainties, some studies still confirmed the existence of the GCE \cite{Zhou:2014lva,Calore:2014xka}.
The origin of the GCE may be astrophysical sources, such as a population of millisecond pulsars (MSP) \cite{Abazajian:2010zy,Yuan:2014rca}. However, the MSP scenario is not easily explained the energy spectrum and spatial distribution of the GCE \cite{Hooper:2013nhl,Cholis:2014noa}.
Additionally, whether emissions from all MSPs can account for the total flux of the GCE is still under debate ~\cite{Yuan:2014rca}.

A very attractive interpretation for the GCE is the DM annihilation. The spatial distribution of the GCE is compatible with the square of Navarro-Frenk-White DM distribution with a slop $\sim \gamma=1.2$. The energy spectrum of the GCE is well fit by a $\sim 30-40$ GeV DM particle annihilating into $b\bar{b}$ with a cross section of $\langle \sigma
v \rangle \simeq 2 \times 10^{-26} \textrm{ cm}^3/\textrm{s}$. A $\sim 7-10$ DM particle annihilating into $\tau \bar{\tau}$ with a cross section of $\langle \sigma v \rangle \simeq 5 \times 10^{-27} \textrm{ cm}^3/\textrm{s}$ is also possible to fit the GCE, but the quality of this fit is lower than the fit for $b\bar{b}$ channel \cite{Daylan:2014rsa}. These models can be directly tested by the Fermi-LAT observations of dwarf spheroidal galaxies. The latest Fermi-LAT results have improved the current limits by a factor of $\mathcal{O}(1)$ and set constraints on the DM models accounting for the GCE \cite{Ackermann:2015zua}. However, considering the uncertainties from the DM density profiles in the GC and dwarf galaxies, the tensions between different Fermi-LAT observations can be relaxed \cite{Calore:2014nla}.

In literatures, simplified models containing new mediators connecting DM particles to $b\bar{b}$ have been proposed to explain the GCE \cite{Agrawal:2014una,Izaguirre:2014vva,Berlin:2014tja}. Since the masses of mediators required by the GCE are within the region of $\mathcal{O}(10^2)-\mathcal{O}(10^3)$~GeV, they can be directly on-shell produced at colliders and then decay into DM particles. At direct detection experiments, even there is no directly interaction between the DM particle and light quarks at tree level, DM-b quark interactions may also induce DM-nucleon scatterings at loop level. Therefore, the results of collider and direct detection experiments would set constraints on simplified models accounting for the GCE \cite{Agrawal:2014una,Izaguirre:2014vva,Berlin:2014tja}.

In principle, UV complete theories can be mapped into simplified models \cite{Cheung:2014lqa}. Supersymmetry (SUSY) is an very attractive and theoretical motivated new physics model; it can provide suitable DM candidates and mediators to explain the GCE. In the SUSY model, DM particles can annihilate into $b\bar{b}$ via t-channel exchange of light sbottoms. The limits on directly sbottom pair production from LEP and LHC results are very stringent. A more promising s-channel annihilation channel is through a pseudoscalar $\chi \bar{\chi} \rightarrow a \rightarrow b\bar{b}$. Comparing with the annihilation mediated by a scalar, this process is not suppressed by the small DM velocity in the Galaxy. Since the pseudoscalar mediator accounting for the GCE is required to be lighter than $\sim 100$ GeV, the LHC Higgs searches have set strong limits on the pseudoscalar in the context of the minimal supersymmetric standard model (MSSM). An economical SUSY extension model evading all collider and direct detection limits is the next-to-minimal supersymmetric standard model (NMSSM) \cite{Ellwanger:2009dp}, which provides a correct SM-like Higgs mass $\sim$125~GeV and solves the $\mu$ problem in the minimal supersymmetric standard model. In the NMSSM, the pseudoscalar mediator is the lightest CP-odd Higgs. As pointed in Ref.~\cite{Cheung:2014lqa}, the GCE and DM relic density can be simultaneously interpreted by the Singlino/Higgsino DM resonant annihilation or the Bino/Higgsino DM off-resonant annihilation via the exchange of light pseudoscalar (see also Refs. \cite{Huang:2014cla,Guo:2014gra,Cao:2014efa,Cahill-Rowley:2014ora}).

Another fundamental problem in the Cosmology is the origin of the baryon asymmetry. An attractive scenario is the electroweak baryogenesis (EWBG), where a strongly first order electroweak phase transition (SFOEWPT) is required to avoid the washout of the generated baryon asymmetry of Universe (BAU)\footnote{In fact, SFOEWPT is not enough for the successful EWBG and the strength of
the SFOEWPT may affect the CP violation source which drives the EWBG~\cite{Jiang:2015cwa}.} . The EWBG connects the Cosmology to the Higgs physics and can be tested at high energy colliders. In the MSSM, a light stop is required to successfully trigger the SFOEWPT~\cite{Huet:1995sh,Carena:1996wj,Carena:2002ss,Carena:2008vj}. After the Higgs discovery, a big challenge is how to simultaneously explain the correct Higgs mass $\sim 125$ GeV  while evading stringent constraints from Higgs measurements. In fact, recent results of LHC Higgs and stop searches have ruled out the mass window of the light stop in the MSSM\cite{Carena:2012np,Curtin:2012aa,Cohen:2011ap,Cohen:2012zza}.
If the MSSM Higgs sector (consisting of two Higgs doublets) is enlarged to include an additional singlet superfield, we have an alternative
supersymmetric framework, namely the next-to-minimal supersymmetric Model (NMSSM), which could provide the
successful SFOEWPT, as been appreciated for a long while~\cite{Pietroni:1992in,Davies:1996qn,Funakubo:2005pu,Huber:2000mg,Balazs:2013cia,Carena:2011jy,Huang:2014ifa,Kozaczuk:2014kva}.
As pointed in Ref. ~\cite{Huang:2014ifa}, the SFOFWPT would occur in $R_{\kappa} \sim -1$ and positive $R_{\kappa} \leq \mathcal{O}(10)$ in NMSSM, where $R_{\kappa}\equiv 4 \kappa v_s/A_\kappa$ is a critical parameter. In this case, the lightest CP-odd Higgs is very light. It is interesting to find the connection between the DM phenomenology and the EWBG. In this paper, we have studied this problem in detail and found the suitable parameter space simultaneously accommodating the GCE, DM relic density and SFOEWPT in the NMSSM.

This paper is organized as follows. In Sec. II, we reviewed and resummarized the key of the strong first order electroweak phase transition in the NMSSM, and we make the energy gap analysis which gives the clue of EWPT. In Sec. III, the interpretations of GCE in NMSSM is reviewed, the resonance effect in the situation of
$\kappa/\lambda \ll 1$ is analyzed. In Sec.IV, the numerical analysis of the EWPT, GCE and DM relic density are carried out, the benchmark scenario which could
explain GCE and give rise to the SFOEWPT and correct DM relic density is presented. In Sec.V, we summarized the work and give our discussions and conclusion.

\section{Strongly First Order Electroweak Phase Transition in the NMSSM}
\label{sec:2}
The NMSSM model can solve the $\mu$ problem in the MSSM and provide a 125~GeV Higgs boson without large loop corrections.
Throughout this paper we restrict ourselves to the
$Z_3$ NMSSM, where an extended superpotential is given by
\begin{equation}
W = \lambda S H_u H_d  + \frac{1}{3}\kappa S^3,
\end{equation}
After the singlet filed $S$ getting a vacuum expectation value (VEV) $v_s$, an effective $\mu$ term can be naturally generated
 \begin{equation}
\mu \equiv \mu_{eff}  =  \lambda v_s,
\end{equation}
which is expected to be of $\mathcal{O}(10^2)$ GeV. The soft breaking terms in the Higgs sector are given by
 \begin{equation}
V_{soft}= \lambda A_\lambda S H_u H_d + \frac{1}{3}
\kappa A_\kappa S^3  + h.c.
\end{equation}
The angle $\beta$ is defined as
\begin{equation}
v_u = v \sin \beta, \;\;\; v_d= v \cos \beta,
\end{equation}
where $v_u$ and $v_d$ are VEVs of $H_u$ and $H_d$ respectively, and $v=\sqrt{v_u^2+v_d^2}=174$ GeV. Compared with the MSSM, the tree-level Higgs mass in the NMSSM is enhanced by a new term $\lambda^2 |H_u^0 H_d^0|^2$. Therefore, a 125~GeV Higgs can be easily achieved in the NMSSM. The mass square of the SM-like Higgs is given by
\begin{equation}
m_h^2=m_Z^2 \cos^2 2\beta+ \lambda^2 v^2 \sin^2 2\beta + \delta m^2_{loop} + \delta m^2_{mix},
\end{equation}
where $\delta m^2_{loop}$ and $\delta m^2_{mix}$ denote the loop effects and the mixing effects on the Higgs mass.

The dynamics of the EWPT are governed by the finite temperature effective Higgs potential, which reads
\begin{equation}
\label{Veff}
V_{\rm eff} = V_{\rm Tree} + V_{\rm CW} + V_{\rm CT} + V_{\rm T} + V_{daisy}.
\end{equation}
In Eq. \ref{Veff}, $V_{\rm Tree}$ denotes the tree-level Higgs potential in the NMSSM.
$V_{\rm CW}$ is the well-known Coleman-Weinberg potential at zero temperature ~\cite{Quiros:1999jp}.
\begin{align}
\label{Vcw}
    & V_{\rm CW} = \sum_i \frac{(-)^{2s_i}n_i}{64\pi^2}m_i^4(\x_l)\left(\ln\frac{m_i^2(\x_l)}{Q^2}
 -\frac{3}{2}\right)
\end{align}
where $i$ runs over all particles in the NMSSM, with each having degrees
of freedom $n_i$, field-dependent mass $m_i(\varphi_l)$ and spin
$s_i$. In this work, we adopt the Landau gauge and the $\overline{\rm DR}$ scheme with renormalization scale Q to calculate $V_{\rm CW}$. In order to maintain tree-level relations involving VEVs, counter terms $V_{\rm CT}$ should be introduced.
$V_{\rm T}$ denotes the thermal correction at the finite temperature,
\begin{equation}\label{VT}
 V_{\rm T} = \frac{T^4}{2\pi^2} \sum_{i} \pm n_i \int_0^\infty dx\;x^2 \ln(1\mp e^{-\sqrt{x^2+m_i^2/T^2}}).
\end{equation}
which can be improved by daisy resummation contributions $V_{daisy}$. In practice, this term can be achieved by inserting thermal mass contributions in the field-dependent mass.

To avoid the baryon asymmetry generated at the EWPT being washed out, the phase transition must be strongly first
order, which can be quantitatively characterized in the perturbative calculation as
\begin{equation}
v_c(T_c)/T_c \gtrsim 0.9.
\end{equation}
Here $T_c$ and $v_c$ are the critical temperature and order parameter of the phase transition.

We implement the effective potential Eq.~\ref{Veff} into the public package {\tt CosmoTransition}~\cite{Wainwright:2011kj}
to evaluate the phase transition numerically and numerically perform a parameter space scan. As pointed out in previous works~\cite{Funakubo:2005pu,Espinosa:2011ax,Huang:2014ifa}, there are mainly three phase transition patterns in the NMSSM. When the temperature decreases, the Universe can directly transit from the symmetry phase into the electroweak breaking vacuum $\Omega_{\rm EW}$ (Type-III), or undergo a intermediate phase in the singlet subspace (Type-I) or the $H_u$ subspace (Type-II). It has been found that a strongly phase transition can be achieved in Type-I and Type-III transition without light squark contributions to thermal loops~\cite{Huang:2014ifa}.

The shape of the effective potential at zero temperature, more exactly, the energy gap between the symmetry phase and the broken phase $(\Delta V \equiv V_{Sym}-V_{EW})\mid_{T=0}$ encodes the information on EWPT in the NMSSM. A smaller $\Delta V $ may lead to a lower $T_c$ and thus a large transition strength $v_c/T_c$. This correlation can be understood by
\begin{equation}
\frac{v_c}{T_c} \sim \left( v\frac{\partial V}{\partial T}\mid_{T=T_c} \right) \frac{1}{\Delta V}
\label{corr}
\end{equation}
Following this insight, we can perform a semi-analytical analysis of the transition strength in terms of the energy gap~\cite{Huang:2014ifa}.

First of all, the potential energy of the electroweak vacuum can be divided into three parts at tree level
\beq
V_{\rm EW} = V_{\rm EW}^H +  V_{\rm EW}^S + V_{\rm EW}^{HS},
\eeqn
where the individual contributions are given by
\begin{align}
V_{\mathrm{EW}}^H &=-\frac{v^2}{4}M_{Z}^2 \left(\cos^22\beta+\frac{\lambda^2}{g^2}\sin^22\beta \right)\simeq -\frac{v^2m_h^2}{4}, \label{VEWH}\\
V_{\mathrm{EW}}^S &= -\frac{1}{3}{\kappa A_\kappa}v_s^3-\kappa^2 v_s^4, \label{VEWS}\\
V_{\mathrm EW}^{HS}&=-\left(1-\frac{A_\lambda}{2\mu}\sin2\beta- \frac{\kappa}{\lambda}\sin2\beta\right)\mu^2v^2 \equiv -C_A\mu^2v^2\; \label{VEWHS}.
\end{align}
Explicitly, the first part is the contribution from the doublet and is almost fixed by the SM-like Higgs mass. The second part is the pure contribution from the singlet. The third part results from the doublet-singlet mixing,
which can be described by an auxiliary parameter $C_A$,
\begin{equation}
C_A \equiv 1- (\frac{A_\lambda}{2 \mu}+\frac{\kappa}{\lambda}) \sin 2\beta .
\end{equation}

In the case of type-I phase transition, the universe transits into an intermediate phase in the singlet subspace before the
EWPT. The absolute minimum of the singlet subspace at zero temperature locates at the origin or
\begin{align}
\label{US}
u_s=\frac{-A_\kappa}{4\kappa}\left( 1+\sqrt{1-8x_\kappa}~\right).
\end{align}
where $x_\kappa$ is given by
\begin{align}
\label{xk}
x_\kappa= \frac{1}{8}-\frac{1}{8}(1+R_\kappa)^2-C_A\lambda^2 v^2/A_\kappa^2,
\end{align}
where $R_\kappa \equiv 4\kappa \mu/\lambda A_{\kappa}$. The Tpyp-I transition occurs only if the $u_s$ is the absolute minimum with $x_\kappa < 1/9$.
The relevant energy gap can be easily obtained as~\cite{Huang:2014ifa}
\begin{align}
\label{VSG}
\Delta V &=V_S-V_{\rm EW}^H -  V_{\rm EW}^S - V_{\rm EW}^{HS}\nonumber\\
& \simeq \frac{v^{2}m_{h}^{2}}{4}-C_{A}\lambda^{2}v^{2}(u_{s}^{2}-v_{s}^{2})+\kappa^{2}(v_{s}^{2}-u_{s}^{2})^{2} \nonumber \\
 & + \frac{1}{3}\kappa A_{\kappa}\left[2u_{s}^{2}(u_{s}-v_{s})+v_{s}(v_{s}^{2}-u_{s}^{2})\right],
\end{align}
where $V_S$ is the tree-level potential energy of $u_s$.
It is obvious to see that a substantial deviation of $u_s$ from $v_s$ is crucial to decrease the energy gap away from the
doublet limit set by the Higgs mass. The numerical results showed that the Type-I phase transition correspond two cases, in which
either $R_\kappa \lesssim 10 $ and $-1< R_\kappa <0 $~\cite{Huang:2014ifa}. For the large $|R_\kappa|$,  the deviation of $u_s$ from $v_s$ would
be small and is disfavored by the SFOEWPT.

Note that at tree level the vacua energy gap $\Delta V$ may be negative.
Fortunately, the Coleman-Weinberg one loop correction would lift $V_S$ and reduce $V_{EW}$, thus guarantee that the EW vacuum $\Omega_{EW}$ is
lower than $\Omega_S$. However, if the metastable vacuum $\Omega_S$ is well below $\Omega_{EW}$, the finite temperature potential could not smooth out a large negative gap when the Universe cools down. In this case, the phase transition is not valid.

Then we consider the Type-III phase transition arising in the case that the origin is the absolute minimum in the singlet subspace or
the origin is metastable.
In this scenario, the tree-level energy gap is simply given by $-V_{\rm EW}$ as
\begin{align}\label{DVT3}
\Delta V &\simeq  \frac{v^{2}m_{h}^{2}}{4}+C_{A}\mu^{2}v^{2}+\kappa^{2}v_s^{4}(\frac{4}{3 R_{\kappa}}+1).
\end{align}
We can see that the singlet part will dominate the energy gap for a large $\mu$. A small energy gap often requires a large negative $A_{\kappa}$ and $-4/3 \lesssim R_{\kappa} <0 $. For a moderate $\mu$, the mixing part becomes important and a negative $C_{A}$ is useful to decrease the gap.

\section{NMSSM interpretations of Galactic Center Excess}
\label{sec:3}
The Galactic center excess( GCE) prefers a $\sim$30-40 GeV DM particle annihilating directly into $b\bar{b}$ with a cross-section about  $\langle \sigma
v \rangle \simeq 2 \times 10^{-26} \textrm{ cm}^3/\textrm{s}$.
In this work, we would like to explain this excess by the annihilation of neutralino pair to $b\bar{b}$ via an s-channel light CP-odd pseudo-scalar in the NMSSM.

The neutralino mass matrix in the NMSSM is written as
\begin{equation}
{\cal M} = \left(
\begin{array}{ccccc}
M_1 & 0 & -\frac{g_1 v_d}{\sqrt{2}} & \frac{g_1 v_u}{\sqrt{2}} & 0 \\
  & M_2 & \frac{g_2 v_d}{\sqrt{2}} & - \frac{g_2 v_u}{\sqrt{2}} &0 \\
& & 0 & -\mu  & -\lambda v_u \\
& & & 0 & -\lambda v_d\\
& & & & 2 \kappa v_s
\end{array}
\right). \label{eq:MN}
\end{equation}
The lightest mass eigenstate of the neutralino is the DM candidate, which is defined as:
\begin{equation}
\chi = N_{11} \tilde{B}+N_{12} \tilde{W}+N_{13} \tilde{H}_d+N_{14}\tilde{H}_u+N_{15}\tilde{S}.
\end{equation}

The Majorana DM annihilation cross section to $b\bar{b}$ with a relative velocity $v_r$ for the interaction $\frac{i}{2}a(y_{a\chi\chi}\bar{\chi}\gamma^5 \chi+ y_{abb}\bar{b}\gamma^5 b)$ is given by
\begin{equation}
\sigma v_r  \simeq \frac{3 }{32 \pi m_\chi^2} \frac{y^2_{a\chi\chi} y^2_{abb} }{(\delta+v_r^2/4)^2+\gamma^2},
\label{DManncs}
\end{equation}
where $\gamma \equiv m_a \Gamma_a/4 m_\chi^2$ is a parameter defined by the pseudoscalar decay width $\Gamma_a$, $\delta$ is a degeneracy parameter defined as
\begin{equation}
\delta=1-\frac{m_{a}^2}{4 m_{\chi}^2} .
\end{equation}
If $\delta$ is not very small, the DM annihilation occur off-resonance and is almost velocity independent. As discussed in Ref.~\cite{Daylan:2014rsa}, the GCE and correct thermal DM relic density can be simultaneously accommodated as long as the product $y_{a \chi\chi}^2 y_{a bb}^2$ is adjusted to an appropriate value. In this case, the main components of the DM would be Bino and Higgsino in the limit $\kappa/\lambda \gg 1$. Since the singlet scalar is heavy and is almost decoupled, the SM-like Higgs is similar to that in the MSSM. The singlet component of the lightest pseudoscalar can suppress its production cross section, hence the stringent limits on the the pseudoscalar mass from the LHC $H/A \rightarrow \tau^+ \tau^-$ searches can be avoided. However, large $\kappa/\lambda$ would enhance the $-V_{EW}^S$ in Eq.~(\ref{VEWS}), which is disfavored by the SFOEWPT. Furthermore, a very small $\lambda$ is difficult to give a correct SM-like Higgs mass $\sim 125$ GeV at tree level. Therefore, in this work we do not consider such parameter space.

In the limit $\kappa / \lambda \ll 1$, the dominant component of the DM may be Singlino. In this case, the correlation between $\kappa /\lambda$ and the DM mass $m_\chi$ in the $Z_3$ NMSSM is
\begin{equation}
\frac{\kappa}{\lambda}=\frac{m_\chi}{2\mu}\left[ 1-\frac{\lambda^2 v^2 (m_\chi-\mu \sin 2\beta)}{m_\chi (m_\chi^2 -\mu^2)}\right]\sim \frac{m_\chi}{2\mu},
\label{chimass}
\end{equation}
where the second equality corresponds to small $m_\chi/\mu$.
The DM would also have non-negligible Higgsino components with a moderate $\mu$ value, which can be described by for small $m_\chi/\mu$

\begin{eqnarray}
\frac{N_{13}}{N_{15}} &\sim& -\frac{\lambda v}{\mu} \cos \beta (1-  \frac{m_\chi }{\mu} \tan\beta ) \label{N1315} \\
\frac{N_{14}}{N_{15}} &\sim& -\frac{\lambda v}{\mu} \sin \beta (1-  \frac{m_\chi }{\mu \tan\beta}) \label{N1415}.
\end{eqnarray}
For the CP-odd Higgs, the absence of tachyon states favors a negative $\kappa A_\kappa$ or a small positive $\kappa A_\kappa$. The constraints from the LHC Higgs researches also compress the allowed parameter space. If the heavier CP-odd Higgs is very heavy which is consistent with the Higgs observations, the lightest CP-odd Higgs mass can be written as
\begin{equation}
m^2_a\simeq \frac{1}{2}\lambda v^2 \sin2\beta (\frac{\lambda A_\lambda}{\mu}+4 \kappa)-3\frac{\mu}{\lambda}\kappa A_\kappa.
\label{amass}
\end{equation}

If $ \lambda v \sin2\beta/2\mu$ is not very large, the main component of the lightest pseudoscalar is singlet. In this case, the coupling $g_{abb}$ is suppressed by a small active pseudoscalar fraction. At the same time, the coupling $g_{a\chi\chi}$ is determined by small $\kappa$ and small Higgsino fraction. This means present DM annihilation cross section should be enhanced by the resonance effect with small $\delta$. From Eq. \ref{DManncs}, we can see that the GCE can be easily explained by adjusting the combination $g_{a\chi\chi} g_{abb}/(\delta^2+\gamma^2)$ in the zero temperature limit. In the early Universe, the DM annihilation occurring close to the resonance is sensitive to the temperature \cite{Griest:1990kh,Gondolo:1990dk}. If $\delta<0$ and $|\delta|$ is not very small, the $\sigma v_r$ tends to have a larger value with a larger $v_r$ as the temperature increases in the early Universe. In this case, the DM relic density would be suppressed. Some non-thermal DM production mechanisms are needed to obtain the correct DM relic density. We do not consider such mechanisms and focuss on the thermal freeze-out mechanism here. For the $\delta>0$ case, from Eq. \ref{DManncs}, we can see that the $\sigma v_r$ always decreases as the temperature increases. This means the process $\chi\chi \rightarrow a \rightarrow b\bar{b}$ can not sufficiently reduce the DM abundance in the freeze-out epoch. Some other DM annihilation channels are needed to generate the correct DM relic density.

\section{Numerical results}

In this section, we perform a numerical analysis to interpret the GCE and SFOEWPT in the NMSSM. We use the packages {\tt NMSSMTools 4.4.0} \cite{Ellwanger:2004xm,Ellwanger:2005dv}, {\tt micrOMEGAs 3.6.9.2} \cite{Belanger:2013oya} and {\tt CosmoTransition} \cite{Wainwright:2011kj} to scan a six-dimensional parameter space ($\lambda$, $\kappa$, $A_\lambda$, $\lambda_\kappa$, $\mu$, $\tan\beta$) with various experimental limits. Since we focus on the EWPT properties affected by the singlet sector, we assume that all the sfermions are heavy and fix soft breaking parameters as $M_{L}=M_{E}=200$ GeV and $M_Q=M_u=M_d=A_t=A_b=A_l=$ 2000 GeV. A benchmark point, which can simultaneously explain the GCE and SFOEWPT, is given in Tab .\ref{DMbm}.

 \begin{table}[h]
 \begin{center}
 \scalebox{1}{
  \begin{tabular}{c|c|c|c|c|c} \hline
  $\lambda$ & $A_\lambda$ (GeV) & $\kappa$ & $A_\kappa$ (GeV) & tan$\beta$ & $\mu$ (GeV)    \\ \hline
   0.5 & 840.0 & 0.029 & -99.18 & 3.05 & 235.0 \\ \hline
$m_{\tilde{\chi}^0_1}$ (GeV)&$m_a$ (GeV) & $\Omega h^2$ &  $\langle \sigma v \rangle|_{v \to 0}(\rm{cm}^3/\rm{s}) $ &  $\sigma_{\rm{SI}} (\rm{cm}^2)$  & PTS  \\ \hline
34.96 & 69.54&0.095 & $1.72\times 10^{-26}$ & $9.03\times 10^{-10}$ & 1.06 \\ \hline
  \end{tabular}}
 \end{center}
\caption{The benchmark point which can simultaneously explain the GCE and SFOEWPT.
\label{DMbm}}
\end{table}

\begin{figure}[!htb]
\includegraphics[width=.45\textwidth]{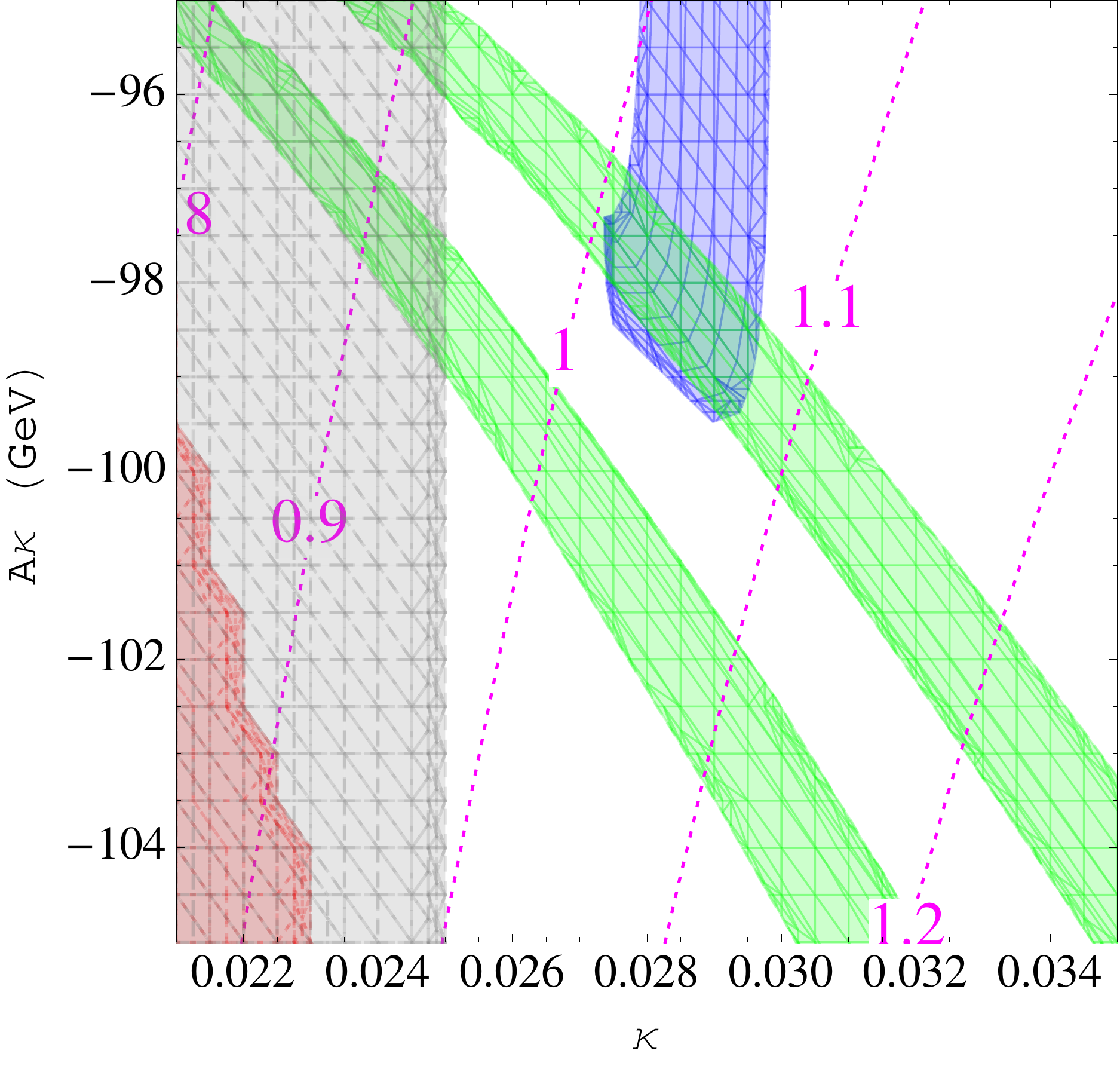}
\includegraphics[width=.45\textwidth]{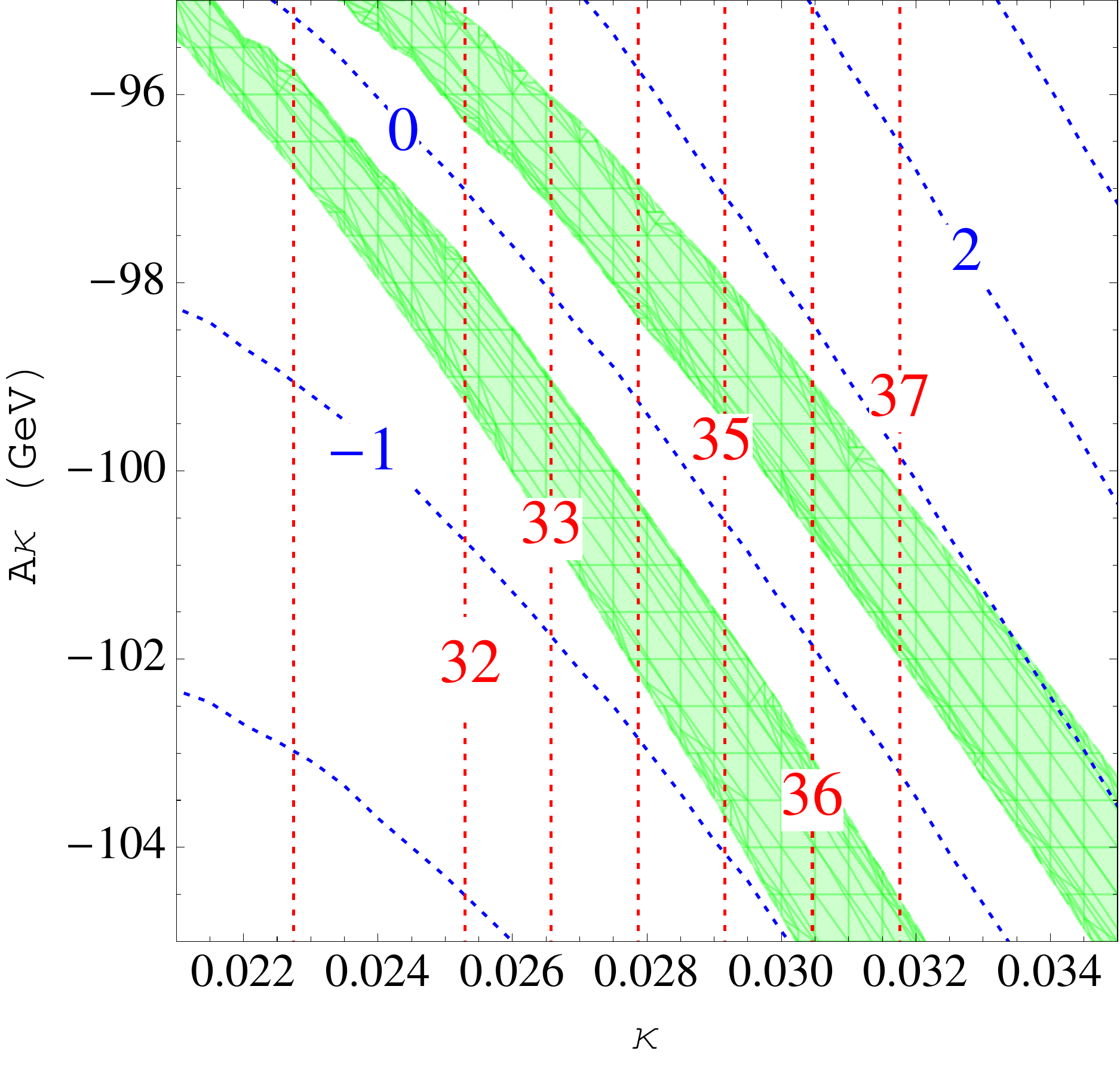}
\caption{Results of a numerical scan with fixed: $\lambda=0.5,~A_\lambda=840$ GeV, $\mu=235$ GeV, $\tan\beta=3.05$, $M_1=M_2/2=100$ GeV, $M_3=800$ GeV, $M_{L}=M_{E}=200$ GeV, $M_Q=M_u=M_d=A_t=A_b=A_l=$ 2000 GeV . The green region denote the parameter space where the DM annihilation
$\sigma v_r \sim (0.5 - 4)\times 10^{-26}$ cm$^3$ s$^{-1}$ can interpret the GCE. In the left panel, The blue region can explain the correct DM relic density $ 0.091< \Omega h^2 <0.138$ \cite{Ade:2013zuv,Hinshaw:2012aka} via the Z mediated annihilation in the early Universe. Also shown are contours of $v_c/T_c$ (the magenta dotted lines) which is usually required to be larger than 0.9 by the SFOEWPT. The red and gray regions are excluded by the direct detection LUX \cite{Akerib:2013tjd} and the Higss mass 124~GeV$< m_h<$ 128~GeV. In the right panel, we show contours of the DM mass $m_\chi$ (red dotted lines) and mass difference $2m_\chi-m_a$ (blue dotted lines).}
\label{kak}
\end{figure}

\begin{figure}[!htp]
\includegraphics[width=.45\textwidth]{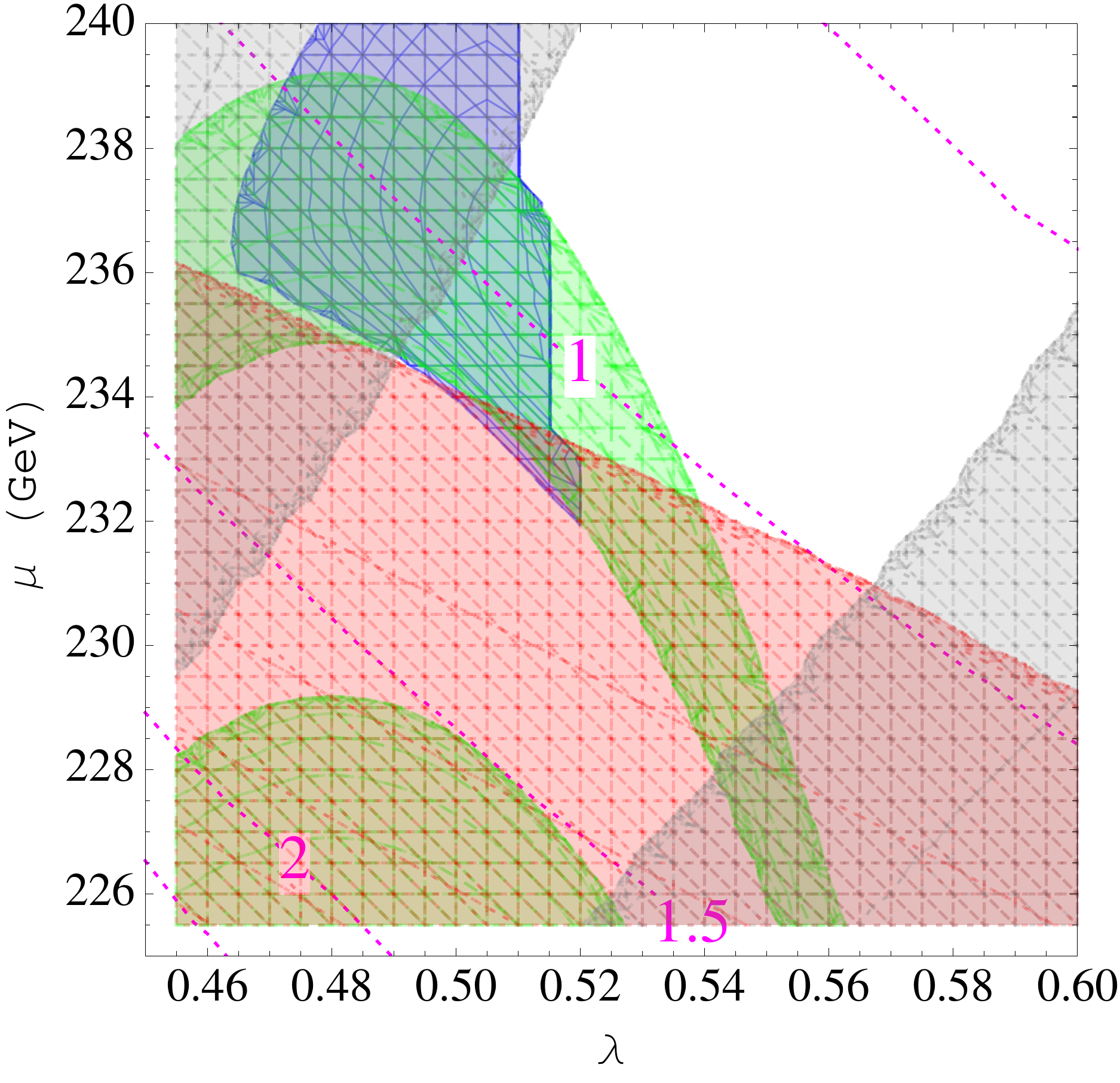}
\includegraphics[width=.45\textwidth]{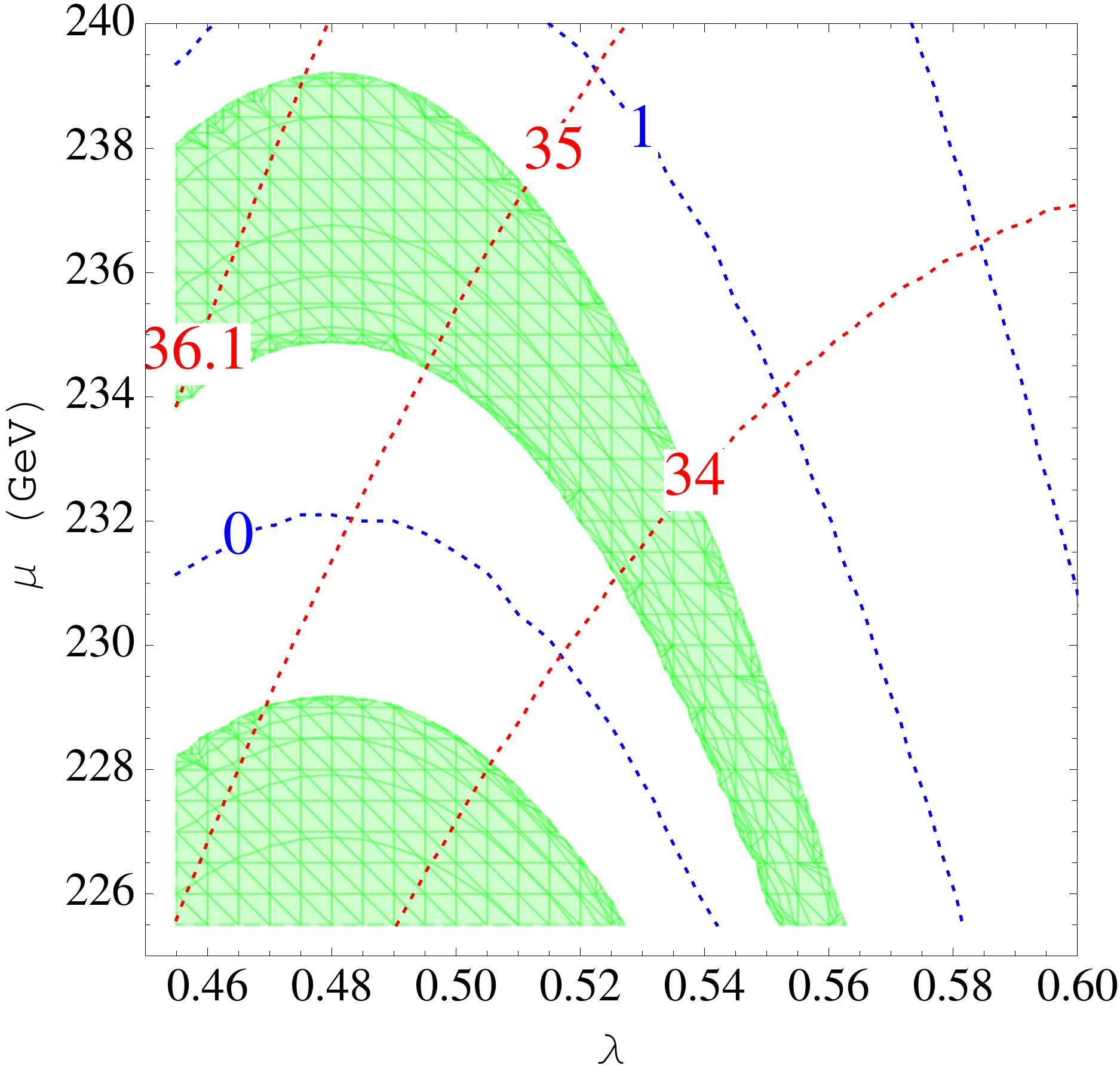}
\caption{Results of a numerical scan with varying $\lambda$ and $\mu$, $\kappa=0.029$, $A_\kappa=-99.18$ GeV and other parameters being fixed as Fig.~\ref{kak}. Other discerption are same as Fig.~\ref{kak}.}
\label{mulam}
\end{figure}

In order to understand the properties of the parameter distribution favored by the GCE and SFOEWPT, we
plot Fig.~\ref{kak}, where only two parameters $\kappa$ and $A_\kappa$ vary in the parameter space near the benchmark point and the other parameters are fixed. In the right panel of Fig .\ref{kak}, red and blue lines denote DM mass $m_\chi$ and mass difference of $2m_\chi-m_a$. The behavior of these lines can be easily understood from Eq. \ref{chimass} and Eq. \ref{amass}. In Fig .\ref{kak}, we show the region consistent with the GCE with an annihilation cross section $\sigma v_r|_{v_r \rightarrow 0}$ in the range of $(0.5 \sim 4)\times 10^{-26}$ cm$^3$ s$^{-1}$. As expected from Eq. \ref{DManncs}, there are two distinct strips near the contour $2m_\chi-m_a=0$ which correspond to the cases $\delta >0$ and $\delta<0$ respectively.

If there is no other DM annihilation channel, it is difficult to explain the GCE and correct DM relic density simultaneously. In the lower green band for $\delta <0$, the DM annihilation occurring closer to the resonance in the early Universe would induce a very small relic density, while in the upper green band for $\delta >0$, the DM particles would be overproduced due to the suppression of the DM annihilation cross section in the early Universe. In the left panel of Fig .\ref{kak}, the blue region denotes the parameter space where a correct DM relic density can be obtained. In this region, the DM annihilation via the exchange of the Z boson plays an important role in the early Universe, and sufficiently reduce the overproduction of DM for the $\delta >0$ case. This annihilation channel is a p-wave process and can be negligible in the Galaxy with small DM relative velocity. The green and blue bands can overlap at $(\kappa, A_\kappa)\sim$ (0.029, -99.2~GeV). Since the correct DM relic density requires a non negligible DM-Z interaction, we also check the partial decay width of the Z to a pair of DM which is controlled by the Higgsino component of DM. For our benchmark point, this width is 0.59~MeV and is allowed by the upper limit $\sim 2$~MeV from the Z decay measurements \cite{ALEPH:2005ab}.

Then we show the parameter distribution of $(\lambda, \mu)$ favored by DM results in Fig .\ref{mulam}, where the other parameters are taken in Tab .\ref{DMbm}. Similar to Fig .\ref{kak}, there are two bands for the GCE correspond to different $\delta$ values. Eqs. \ref{N1315} and \ref{N1415} imply that the Higgsino faction of the DM and thus the DM coupling to the Z boson are controlled by $\lambda/\mu$. By tuning the value of $\lambda/\mu$, the correct DM relic density can be obtained via the Z channel annihilation in the blue region. The upper green band and blue region overlap at $(\lambda, \mu)\sim$ (0.50, 235~GeV).

In Fig. \ref{kak} and \ref{mulam}, we also show the constraint from the direct detection. The LUX collaboration has set stringent constraints on the DM with mass $\sim \mathcal{O}(10)$GeV \cite{Akerib:2013tjd}. The spin-independent signature for our benchmark point is dominantly induced by the DM-quark scattering process via the exchange of CP-even Higgs. The scattering cross section is strongly affected by Higgsino components in the DM. This explains why the parameter space with large $\lambda$ and small $\mu$ would be excluded by the LUX results.

\begin{figure}[!htb]
\begin{minipage}[t]{0.45\textwidth}
    \centering
    \includegraphics[scale=0.4]{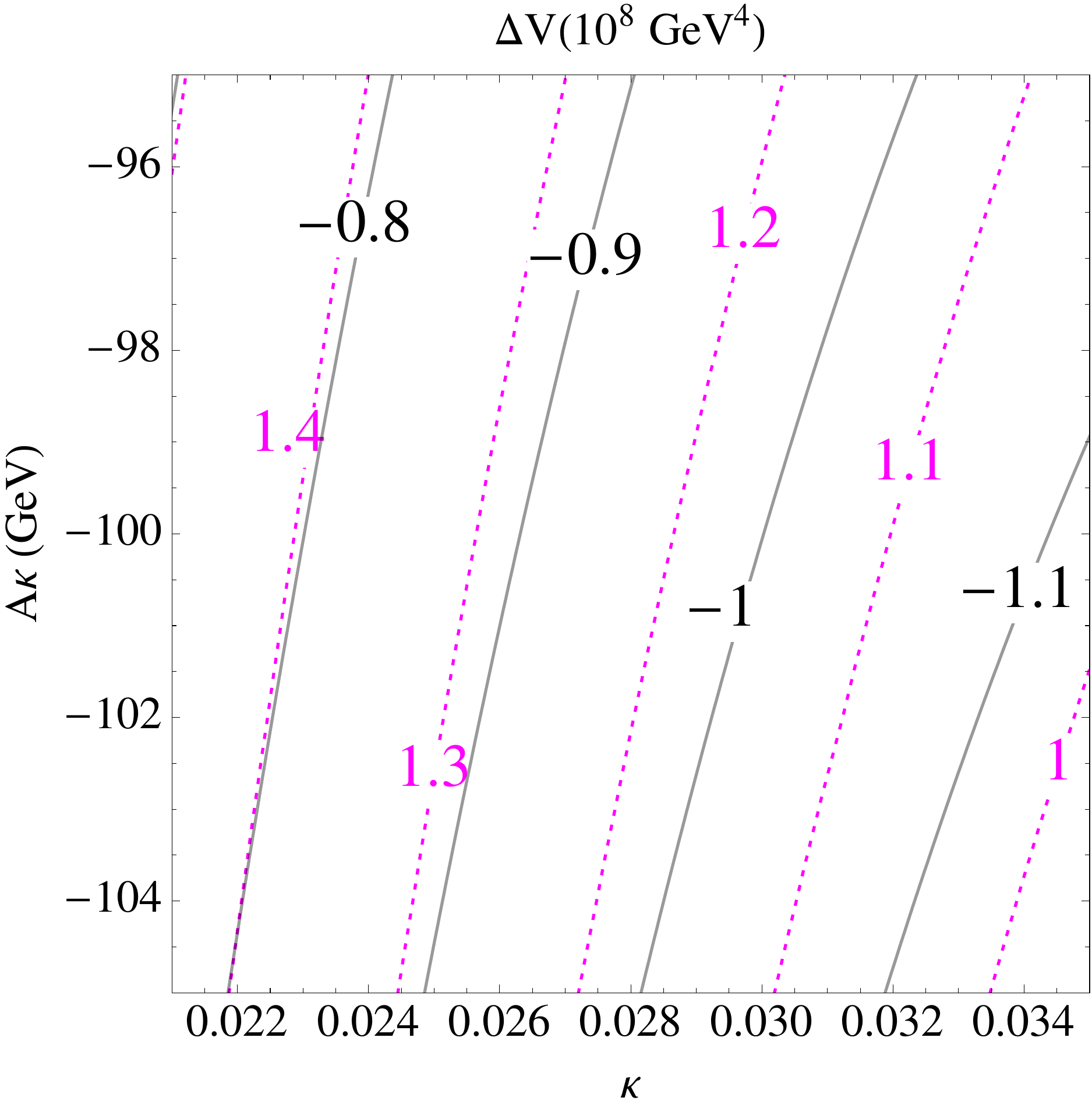}
\end{minipage}
\hspace{0.5cm}
\begin{minipage}[t]{0.45\textwidth}
    \centering
    \includegraphics[scale=0.4]{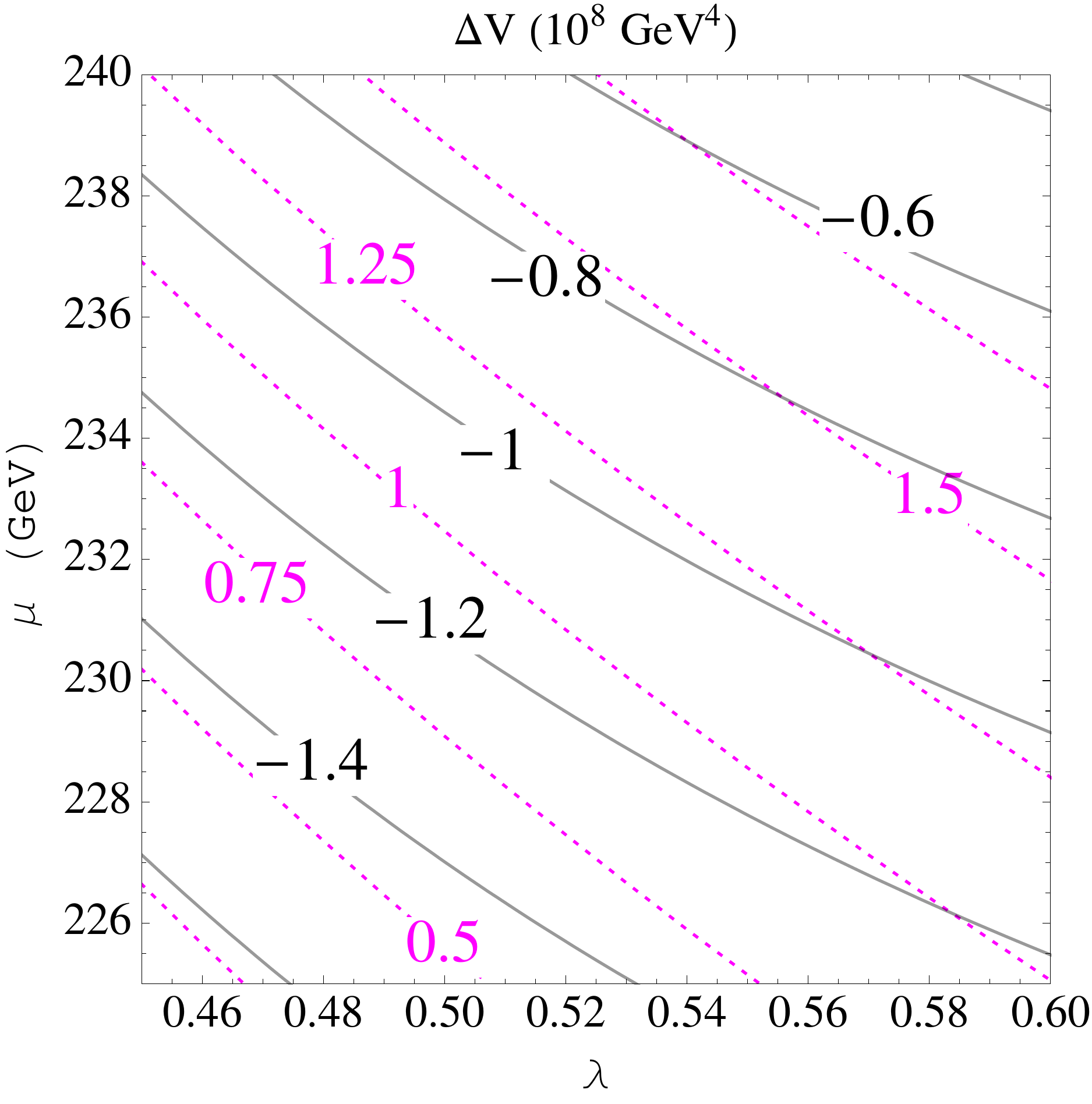}
\end{minipage}
\caption{The energy gaps at tree-level $\Delta V_{tree}$ ( black contours) and the full numerical results with loop corrections $\Delta V_{ful}$  (magenta contours). Left panel: plots of $\Delta V$ in the $\kappa$-$A_\kappa$ plane with
parameters set as those in Fig.~\ref{kak}; Right panel: plots of $\Delta V$ in the $\lambda$-$\mu$ plane with parameters set as those in Fig.~\ref{mulam}.}
\label{delt_V}
\end{figure}

We depict the contours of $v_c/T_c$ in Figs .\ref{kak} and \ref{mulam}. It is can be found that there does exist a parameter space satisfying the DM results and SFOEWPT.
Eq.~\ref{corr} implicates the relation between the phase transition strength and the energy gap, deeper understanding of the SFOEWPT calls for a detail analysis of
the energy gap. The energy gap at tree level $\Delta V_{tree}$ and full numerical result with the Coleman-Weinberg one loop correction $\Delta V_{ful}$ are shown in Fig. \ref{delt_V}. We find that the variation tendency of phase transition strength $v_c/T_c$ with respect to $\lambda$($\kappa$) and $\mu$($A_\kappa$) is consistent with those of the energy gaps $\Delta V_{tree}$ and $\Delta V_{ful}$.

In our benchmark scenario, Type III SFOEWPT is preferred since the mixing term with a negative $C_A$ tends to sufficiently decrease the energy gap. The parameter dependence of the $\Delta V_{tree}$ can be easily understood by Eq. \ref{DVT3} which is rewritten as
\begin{align}
\label{DVtype3}
\Delta V &\simeq  \frac{v^{2}m_{h}^{2}}{4}-\frac{1}{2}\sin 2\beta A_\lambda \mu v^{2}+(1-\frac{\kappa}{\lambda}\sin 2\beta)\mu^{2} v^{2}+\frac{1}{3}\kappa A_\kappa (\frac{\mu}{\lambda})^3+ \kappa^2(\frac{\mu}{\lambda})^4 ,
\end{align}
where the first term, middle two terms, and last two terms represent the contributions from the doublet, mixing, and singlet parts, respectively.
For our benchmark model with $\kappa \sim \mathcal{O}(0.01)$ and $A_\kappa \sim -\mathcal{O}(100)$~GeV, a negative contribution of $\kappa A_\kappa v_s^3/3$ dominates in the singlet part. Hence, $\Delta V_{tree}$ decreases as $\kappa$ and $|A_\kappa|$ increase in the region of $\kappa < -A_\kappa/6 v_s \sim 0.035$, and then the phase transition is enhanced as shown in Fig.~\ref{delt_V} and Fig.~\ref{kak}. From Eq. \ref{DVtype3}, we also find that $\Delta V_{tree}$ is suppressed by the small $\lambda$ in the mixing and singlet parts. The dependence of $\Delta V_{tree}$ on $\mu$ is more complicated. In our benchmark scenario with a small $\kappa$, the contribution from the mixing part would be more important than that from the singlet part. In this case, it can be found that $\Delta V_{tree}$ becomes more negative as $\mu$ decreases when $\mu > A_\lambda \sin 2\beta /4 \sim 110$ GeV.

Note that although the $\Delta V$ at tree level is negative in Fig.~\ref{delt_V}, the loop corrections from the Coleman-Weinberg potential lifts up the $\Delta V_{tree}$ and helps the phase transition to occur in the early Universe. The detailed discussions of the loop corrections to $\Delta V$ can be found in Ref. \cite{Huang:2014ifa}.

\section{Conclusions and discussions}

In this work, we study the possibility that the GCE and SFOEWPT can be simultaneously explained in the NMSSM model.
The GCE can be interpreted by the annihilation process $\chi \chi \rightarrow b \bar{b}$ via the exchange of a light pseudoscalar.
For the Singlino/Higssino DM, the annihilation should occur near the resonance with a very small mass difference $2m_\chi-m_a >0$.
In this case, the correct DM relic density is obtained by the DM annihilation via the exchange of a Z boson in the early Universe, meanwhile
the present DM annihilation cross section accounting for the GCE can also be easily achieved.

Besides the mostly singlet pseudoscalar $\sim \mathcal{O}(10)$~GeV required by the GCE, there is also an accompanying light CP-even Higgs in the scalar sector.
Such singlet like scalar could decrease the vacua energy gap $\Delta V$ with a negative $A_\kappa$ and a moderate $\mu$, which would lead to SFOEWPT required by successful EWBG.
We find that there does exit the parameter space in which the GCE and the correct DM relic density can be interpreted by the Singlino/Higssino DM with
SFOEWPT being realized at the same time.

The discussions in this work can be extended to the DM models containing new singlet states. In such models, the Higgs spectra would be affected
by the requirements from the EWPT and DM phenomenology. Discovering such new scalar sector at $\mathcal{O}(10)$GeV is an excellent motivation for the future high energy collider experiments.

\begin{acknowledgments}
This work is supported by the National Natural Science Foundation of China under Grants NO. 11475189, 11475191, 11135009, 11175251, and the 973 Program of China under Grant No. 2013CB837000
\end{acknowledgments}

\end{document}